\documentclass[aps, prb, 10pt, twocolumn,floatfix,superscriptaddress,longbibliography, nofootinbib]{revtex4-2}

\usepackage[english]{babel}

\usepackage{braket}
\usepackage{amsmath}
\usepackage{graphicx}
\usepackage{amssymb}
\usepackage[colorlinks=true, allcolors=blue]{hyperref}
\usepackage{caption}
\usepackage{subcaption}
\usepackage{tikz}
\usetikzlibrary{quantikz2}  % For older distributions, use "quantik z2"instead

\newcommand{\eql}[2]{\begin{equation} \label{eq:#1} #2\end{equation}}

\newcommand{\eqn}[2]{\begin{equation*} \label{eq:#1} #2\end{equation*}}
\newcommand{\al}[2]{\begin{align} \label{eq:#1} #2\end{align}}
\newcommand{\aln}[2]{\begin{align*} \label{eq:#1} #2\end{align*}}

\begin{document}
% \maketitle

\title{Quantum speed-up for solving the one-dimensional Hubbard model using quantum annealing}
\author{Kunal Vyas}
\affiliation{Jülich Supercomputing Centre, Forschungszentrum Jülich, D-52425 Jülich, Germany}
\affiliation{RWTH Aachen University, 52056 Aachen, Germany}
\author{Fengping Jin}
\affiliation{Jülich Supercomputing Centre, Forschungszentrum Jülich, D-52425 Jülich, Germany}
\author{Hans De Raedt}
\affiliation{Jülich Supercomputing Centre, Forschungszentrum Jülich, D-52425 Jülich, Germany}
\author{Kristel Michielsen}
\affiliation{Jülich Supercomputing Centre, Forschungszentrum Jülich, D-52425 Jülich, Germany}
\affiliation{Department of Computer Science, University of Cologne, 50931 Cologne, Germany}

\begin{abstract}
The Hubbard model has occupied the minds of condensed matter physicists for most part of the last century. This model provides insight into a range of phenomena in correlated electron systems. We wish to examine the paradigm of quantum algorithms for solving such many-body problems. The focus of our current work is on the one-dimensional model which is integrable, meaning that there exist analytical results for determining its ground state. In particular, we demonstrate how to perform a gate-based quantum computer simulation of quantum annealing for the Hubbard Hamiltonian. We perform simulations for systems with up to 40 qubits to study the scaling of required annealing time for obtaining the ground state. We find that for the half-filled cases considered, there is a substantial quantum speed-up over algorithms based on the Bethe-ansatz equations.

\end{abstract}

\maketitle

% \begin{multicols}{2}
\section{Introduction}\label{INTRO}

The Hubbard model is believed to be one of the simplest models that can capture qualitative behaviour of correlated fermions which cannot be reconciled within band theory \cite{hubbard_electron_1963}. The Hamiltonian for a one-dimensional lattice with open boundaries emerging from this model is given by
\begin{align}H =& -t_H \sum\limits_{\substack{i=1\\ \sigma=\uparrow,\downarrow}} ^{L-1}(c_{i\sigma}^\dagger c_{i+1\sigma}^{\phantom{\dagger}} + c_{i+1\sigma}^\dagger c_{i\sigma}^{\phantom{\dagger}})
\nonumber \\
&+ U \sum\limits_{i=1}^L n_{i\uparrow}n_{i\downarrow}
\;,
\label{eq:hh}
\end{align}
where $L$ is the number of sites of the lattice, $c_{i\sigma}^\dagger(c_{i\sigma}^{\phantom{\dagger}})$ is the fermion creation(annihilation) operator of spin $\sigma$ at site $i$ and $n_{i\sigma} = c_{i\sigma}^\dagger c_{i\sigma}^{\phantom{\dagger}}$ is the number operator of spin $\sigma$ at site $i$. The first sum of terms describes the nearest-neighbor hopping of the electrons, $t_H$ being the hopping integral.
The second sum describes the on-site repulsive interaction with strength $U>0$.

Despite the apparent simplicity of the Hamiltonian, exact results for the solution to the eigenvalue problem of this Hamiltonian only exist in the thermodynamic limit for the one-dimensional lattice \cite{essler_one-dimensional_2005}. Even for the one-dimensional model, the procedure is non-trivial and was first performed by Lieb and Wu \cite{lieb_absence_1968} for periodic boundary conditions. They used the Bethe-ansatz \cite{bethe_zur_1931} to reduce the eigenvalue problem to the Bethe-ansatz equations, which are a set of coupled non-linear algebraic equations, now called the Lieb-Wu equations. The Lieb-Wu equations can be analysed in the thermodynamic limit, and certain properties like the ground state energy can be obtained in a closed form by solving a set of coupled integral equations. The one-dimensional Hubbard model is thus integrable \cite{andrei_integrable_1995} and is considered to be \textit{solved} in this sense.

In the present paper, we study the finite one-dimensional Hubbard model with open boundaries from the perspective of quantum computing. There exist Bethe-ansatz equations also for this version of the problem, which were first calculated by H. Schulz \cite{schulz_hubbard_1985}. The equations for the quantum numbers $\{ k_1,k_2 \dots k_N \}$ and $\{ \lambda_1, \lambda_2 \dots \lambda_{N_\downarrow} \}$ that determine the ground state read
\eql{taka}{2k_j(L+1) = 2\pi j + \sum\limits_{b=\pm 1} \sum\limits_{r=1}^{N_\downarrow} \phi (2 \sin k_j + 2b\lambda_r)\;,}
\al{taka2}{\begin{split}\sum\limits_{b=\pm 1} \sum\limits_{l=1}^{N} \phi (2b &\sin k_l + 2\lambda_r) = \\ &-2\pi r + \sum\limits_{b=\pm 1} \sum\limits_{s\neq 1}^{N_\downarrow} \phi (\lambda_r + b\lambda_s)\;,\end{split}}
where $\phi (x) = -2 \arctan (2xt_H/U)$, $N_\sigma=\sum_{i=1}^{L}n_{i\sigma}$ is the total number of fermions with spin $\sigma$, $N=N_\uparrow+N_\downarrow$ is the total number of fermions, and the ground state energy is given by
\eql{energy}{E = -2t_H \sum\limits_{j=1}^N \cos k_j\;.}

% The ground state is determined by the roots of Eqs.~(\ref{eq:taka}) and (\ref{eq:taka2}) for
% \eqn{int}{I_j = j \quad \text{and} \quad J_r = r,}and 

Going to the thermodynamic limit by taking $L\rightarrow\infty$, and keeping the ratios $N/L$ and $N_\downarrow/N$ constant, Eqs.~\eqref{eq:taka} and \eqref{eq:taka2} turn into coupled integral equations \cite{schulz_hubbard_1985}. These equations can be solved analytically for the case of $2N_\downarrow=N=L$, and the corresponding ground state energy can be calculated. However, for a general $L$, $N$ and $N_\downarrow$, a computer is needed to solve the respective Bethe-ansatz equations. The time required to solve the equations by a root finding algorithm scales polynomially with $L$ and thus, the ground state energy can also be found for large $L$. On the other hand, finding the ground state using the roots is still exponentially expensive due to the requirement of calculating an exponentially large number of coefficients.

For the Grover algorithm \cite{grover_quantum_1997}, it was shown that a quadratic speed-up can be achieved compared to the best-known classical algorithm for searching for an item in a list. This result played a significant role in demonstrating the potential of quantum computers. Similarly, a question emerges in the context of integrable quantum many-body systems, which are polynomially hard to solve: Whether a quantum algorithm can provide any kind of benefit in finding the ground state of such systems. The one-dimensional Hubbard model serves as the perfect toy problem in taking a step towards answering this question. 

Various techniques to address this question are currently being investigated. One of the popular algorithms to obtain the ground state is VQE \cite{peruzzo_variational_2014}. In references \cite{cade_strategies_2020, wecker_progress_2015}, strategies for implementing VQE on a quantum computer are presented. Even though they show good success probabilities for achieving the ground state of Hubbard systems, the scalability of these methods is still unclear.

An alternative way to approach the problem of finding the eigenstates of Eq.~\eqref{eq:hh} is to use a quantum algorithm to prepare the Bethe-ansatz states directly on a quantum computer \cite{van_dyke_preparing_2021, li_bethe_2022}. These papers demonstrate methods to obtain eigenstates of $XX$ and $XXZ$ models using the respective Bethe-ansatz equations. They were able to achieve linear scaling of resources for finding particular eigenstates of these models. However, success probabilities for large systems were not shown.
 
A distinct method that is gaining popularity is the algorithm of quantum annealing \cite{kadowaki_quantum_1998, farhi_quantum_2001, farhi_quantum_2000}, which can be used to find the ground state of a Hamiltonian. Quantum annealing is being investigated prominently for solving optimization problems. Various problems have been approached through such ideas \cite{farhi_quantum_2001, martonak_quantum_2004, brooke_quantum_1999, farhi_quantum_2000}, and there has been some progress in understanding what kind of an advantage they can bring, if any, over classical methods of solving them \cite{munoz-bauza_scaling_2025, somma_quantum_2012, albash_demonstration_2018, mehta_quantum_2022, hsu_quantum_2019, farhi_quantum_2019}. 

On the other hand, one of the most compelling promises of quantum computers is the simulation of quantum systems \cite{feynman_simulating_1982,smith_simulating_2019}. Due to the advent of analog quantum devices, the dynamics of a variety of many-body quantum systems can be implemented on them \cite{gross_quantum_2017, bernien_probing_2017, harris_phase_2018}. Digital quantum computers also offer prospects for modelling strongly correlated quantum systems \cite{wecker_solving_2015,jiang_quantum_2018,smith_simulating_2019, zhang_digital_2012} that are too large to be treated by conventional computers. Moreover, simulating quantum dynamics of these systems on a quantum computer enables the investigation of several physical phenomena, such as phase transitions \cite{wecker_solving_2015, harris_phase_2018}.

In light of all this, we wish to examine the potential of quantum annealing in order to understand its utility for finding the ground state of quantum systems by simulating quantum dynamics of the Hubbard model on a quantum computer. We implement the protocol of quantum annealing on a quantum computer simulator, demonstrating the resource requirement for a faithful gate-based simulation of a quantum system while studying its performance for achieving the ground state of the Hubbard model described by the Hamiltonian Eq.~\eqref{eq:hh}.

We investigate the scaling of the performance of the quantum annealing algorithm to solve the problem under consideration with respect to the system size. We find that the time required to find the ground state of the Hubbard model on a one-dimensional lattice in a subspace with $2N_\downarrow=N=L$, scales at most linearly with system size. This suggests a substantial quantum speed-up over other algorithms to calculate the ground state of the Hubbard model.

We begin by providing the background about adiabatic quantum computing and quantum annealing in section \ref{aqc}. In section \ref{gate-based}, we show how the algorithm of quantum annealing for the Hubbard model can be implemented on a gate-based quantum computer. Subsequently, we present our results of the simulations we perform for systems with up to 40 qubits in section \ref{results} and demonstrate the analysis. Finally, we discuss the conclusions and implications of these results in section \ref{conclusion}.

\section{quantum annealing} \label{aqc}

\subsection{Theory}

Quantum annealing is a method that is based on the exploitation of the quantum adiabatic theorem \cite{born_beweis_1928, sakurai_modern_2020} to find eigenstates of Hamiltonians that are computationally hard to find. The theorem points us towards an idea which is now quite well known that, if the system is in an eigenstate and the Hamiltonian is changed adiabatically, then the system remains in the corresponding eigenstate of the evolving Hamiltonian, provided that the energy gap of the eigenstate with respect to other instantaneous eigenstates does not close during the evolution. This property can be used to calculate the ground state of a Hamiltonian without explicitly solving its eigenvalue equation, which is generally hard if the Hilbert space is large.

Suppose we have a Hamiltonian $H_p$ for which we want to find the ground state given by,
\eqn{ev}{H_p \ket{\phi_{p,0}} = E_{p,0}\ket{\phi_{p,0}}\;,}
where $E_{p,0}$ is the lowest eigenvalue of $H_p$ and $\ket{\phi_{p,0}}$ is the corresponding eigenvector. In lieu of solving this eigenvalue problem, we can prepare a system in a quantum state $\ket{\psi}=\ket{\phi_{i,0}}$, where $\ket{\phi_{i,0}}$ is a known and easy-to-prepare ground state of some initial Hamiltonian $H_i$, and transform the system such that the instantaneous Hamiltonian, say $H(s)$, can be written as,
\eqn{hs}{H(s) = f_i(s)H_i + f_p(s)H_p\;,}where $f_i(s)$ and $f_p(s)$ are real functions of $s$ so that $H(s)$ goes from $H_i \rightarrow H_p$ as $s$ goes from $0 \rightarrow 1$. Provided that the ground state energy of $H(s)$ is always separated by a gap from the energy of the first excited state during this process and that $H(s)$ evolves adiabatically into $H_p$, the state of the system at $s=1$ would be $\ket{\phi_{p,0}}$. This protocol of transforming the Hamiltonian using a single parameter $s$ so that we obtain the ground state of $H_p$ is nowadays called quantum annealing.

Evolution of real systems can never be described by an adiabatically evolving quantum state, which begs the question of how slowly the Hamiltonian should evolve in order to obtain a reasonable estimate of the ground state. If we call $\ket{\phi_j(s)}$ as the $jth$ eigenstate of $H(s)$ and $\ket{\psi(s)}$ as the instantaneous state of the system, the following bound \cite{jansen_bounds_2007} 
\begin{widetext}
\eql{bound}{1-|\braket{\phi_0(1)|\psi(1)}|^2 \leq \frac{1}{T_A^2}\left(\frac{\|\dot{H}(0)\|}{\Delta^2(0)} + \frac{\|\dot{H}(1)\|}{\Delta^2(1)} + \int_0^1 \bigg( 7 \frac{\|\dot{H}\|^2}{\Delta(s)^3} +  \frac{\|\ddot{H}\|}{\Delta(s)^2} \bigg) ds \right)^2
\;,
}
\end{widetext}
serves as a statement of the adiabatic theorem, giving us a window to look at and analyse the behaviour of total quantum annealing time required with respect to system size.
In Eq.~(\ref{eq:bound}), $T_A$ is the total annealing time, $\Delta(s) = E_1(s)-E_0(s)$, and $||A||$ represents the spectral norm of matrix $A$. 

Equation~(\ref{eq:bound}) shows that for a given problem, the transition probability (left side of Eq.~(\ref{eq:bound})) will always be bounded by $T_A^{-2}$ times a value that depends on the gaps in the spectrum and the annealing schedule. Thus, this gap-dependent factor could help us understand how $T_A$ scales across system sizes for a given annealing schedule. 

\subsection{Annealing the Hubbard Hamiltonian}

We investigate the protocol of quantum annealing for a particular schedule to find the ground state of Eq.~\eqref{eq:hh} with half-filled lattices. We want to study how the total annealing time scales to get a reasonable estimate of the ground state with system size. To this end, we simulate quantum annealing for the Hubbard Hamiltonian for a range of system sizes and evaluate the performance. The annealing schedule that we use is,
\eql{schedule}{H(s) = (1-s)H_I + sH_H\;,}
where $H_I$ is the initial Hamiltonian. 

The instantaneous Hamiltonian linearly depends on the anneal parameter $s$, starting from the initial Hamiltonian and ending as the Hubbard Hamiltonian at the end of the anneal. It is important to choose the initial Hamiltonian, such that its ground state is easy to prepare. A natural choice is,
\eql{hd}{H_I = -t_H \sum\limits_{\substack{i=1\\ \sigma=\uparrow,\downarrow}} ^{L-1}(c_{i\sigma}^\dagger c_{i+1\sigma}^{\phantom{\dagger}} + c_{i+1\sigma}^\dagger c_{i\sigma}^{\phantom{\dagger}})\;.}
Clearly, Eq.~(\ref{eq:hd}) is just the hopping part of Eq.~\eqref{eq:hh}. Thus, the instantaneous Hamiltonian for a Hubbard system in one dimension with nearest neighbour hopping and open boundaries can be written as,
\al{ha}{\begin{split}H(s) = -t_H \sum\limits_{\substack{i=1\\ \sigma=\uparrow,\downarrow}} ^{L-1}&(c_{i\sigma}^\dagger c_{i+1\sigma}^{\phantom{\dagger}} + c_{i+1\sigma}^\dagger c_{i\sigma}^{\phantom{\dagger}}) \\ &+ sU \sum\limits_i^L n_{i\uparrow}n_{i\downarrow}\;.\end{split}}

The annealing Hamiltonian evolves from the Hamiltonian for free fermions hopping on a lattice at $s=0$, and the interaction grows in strength as $s$ goes to 1. This implies that for non-zero $s$, the instantaneous Hamiltonian is a Hubbard Hamiltonian with $sU$ as the interaction strength. Hence, this choice of the initial Hamiltonian also helps us to understand how quantum annealing performs for various interaction strengths.

\section{Gate-based simulations} \label{gate-based}

Simulating ideal quantum annealing means solving the time-dependent Schrödinger equation (TDSE) \eql{tdse}{i\hbar \frac{d}{dt}\ket{\psi(t)} = H(t)\ket{\psi(t)}\;,}
for the Hamiltonian Eq.~\eqref{eq:ha} with $t=sT_A$.

Since we are interested in investigating the performance of a quantum computer for finding the ground state using quantum annealing, we simulate Schrödinger dynamics using quantum circuits. To perform gate-based simulations of quantum annealing for Hubbard-like models, we need two kinds of circuits:
\begin{enumerate}
  \item Initial state preparation circuit
  \item Circuit for performing dynamics of quantum annealing
\end{enumerate}

In the following sections, we describe how we construct these circuits to perform quantum annealing for fermion systems on a digital quantum computer.

\subsection{Jordan-Wigner transformation} \label{jwt}

The first step is to transform the fermion Hamiltonian into a qubit Hamiltonian comprising Pauli matrices. To preserve the fermion character of the Hamiltonian, the operators in the transformed Hamiltonian need to obey the same anti-commutation relations as fermion operators. We also need to map the fermion states to qubit states. A straightforward way to do this is to use the Jordan-Wigner transformation \cite{jordan_uber_1928} in which the $2L$ spin-orbitals on $L$ lattice sites are mapped onto $2L$ qubits where each occupied spin-orbital corresponds to $\ket{1}$ and each unoccupied spin-orbital corresponds to $\ket{0}$.

According to the Jordan-Wigner transformation, the fermion creation and annihilation read
\begin{align}
\begin{split} \label{eq:jw}
    &c_i^\dagger \rightarrow \frac{1}{2}Z_1 \dots Z_{i-1} (X_i - iY_i)\;, \\
    &c_i \rightarrow \frac{1}{2}Z_1 \dots Z_{i-1} (X_i + iY_i)\;,
\end{split}
\end{align}
where 
% $c_i^\dagger(c_i)$ creates(annihilates) a fermion at the $i$th spin-orbital and 
$X_i, Y_i, Z_i$ are Pauli matrices acting on the $i$th qubit.

Representing the spin-$\uparrow$ orbitals on positions 1 to $L$ and spin-$\downarrow$ orbitals on positions $L+1$ to $2L$, the transformed Hamiltonian reads
\al{q-ham}{\begin{split}H(s) = -\frac{t_H}{2} \sum\limits_{\substack{i=1 \\ i \neq L}}^{2L-1} &(X_iX_{i+1}+Y_iY_{i+1}) \\ &+ \frac{sU}{4}\sum\limits_{i=1}^L (I-Z_i)(I-Z_{i+L})\;.\end{split}}

Simulating the real-time dynamics of the fermion Hamiltonian Eq.~\eqref{eq:ha} is equivalent to simulating the real-time dynamics of the qubit Hamiltonian Eq.~\eqref{eq:q-ham}. We use the latter to construct quantum circuits that perform quantum annealing and use them to study their performance.

\subsection{Initial state preparation} \label{initial state}

An essential step in performing quantum annealing is to prepare the system in the ground state of the initial Hamiltonian. Therefore, we need a quantum circuit that initializes the state of a quantum computer in the ground state of $H_I$.
% \eql{h0}{H(0) = -t_H \sum\limits_{\substack{i=1\\ \sigma=\uparrow,\downarrow}} ^{L-1}(c_{i\sigma}^\dagger c_{i+1\sigma} + c_{i+1\sigma}^\dagger c_{i\sigma}).}

\begin{figure}
\centering
\includegraphics[width=\columnwidth]{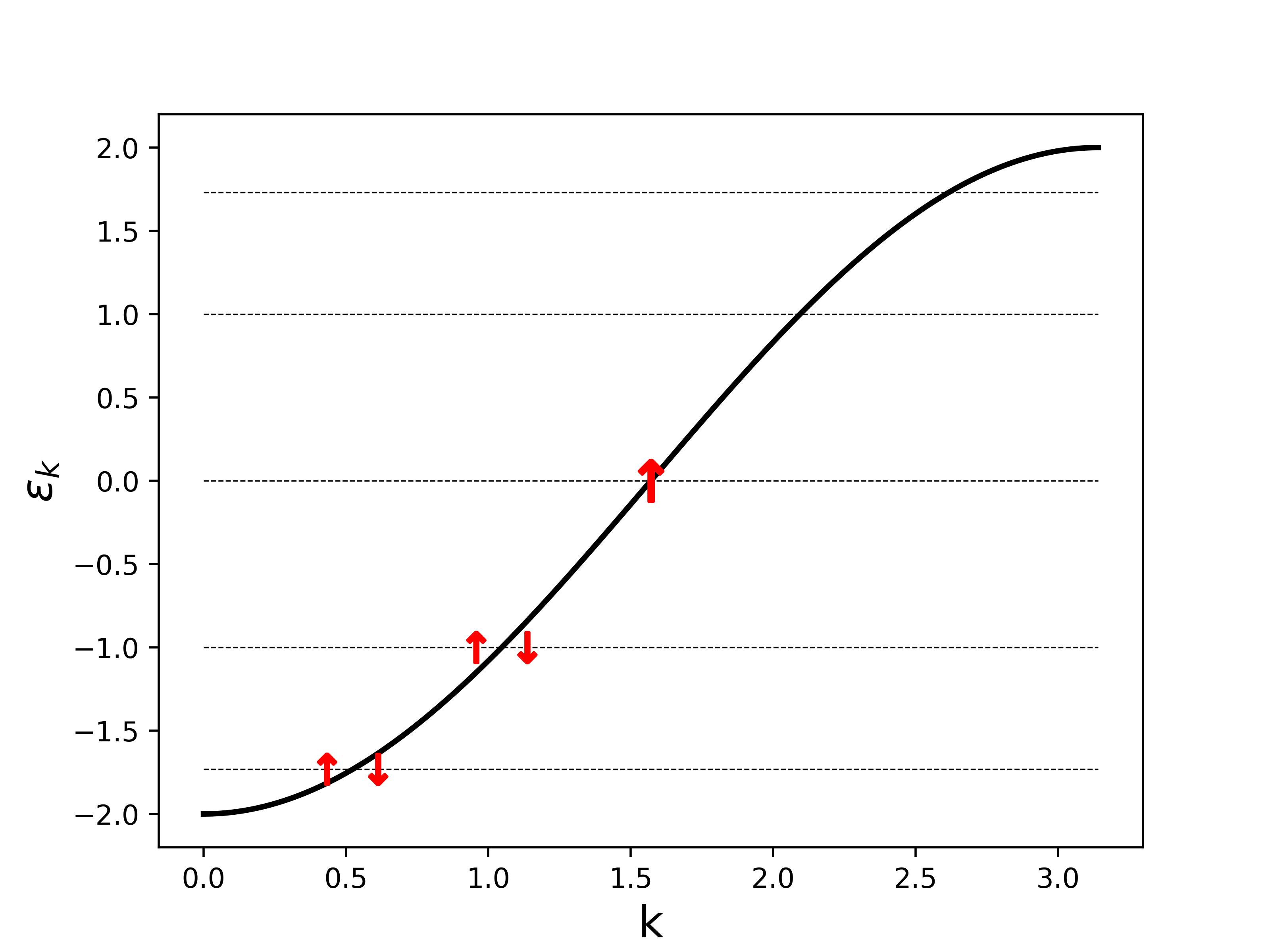}
\caption{\label{fig:dispersion}Dispersion relation and the ground state occupation of the energy levels of the initial Hamiltonian for $L=5$, $N_\uparrow=3$, $N_\downarrow=2$}
\end{figure}

Preparing the ground state of $H_I$ on a quantum computer might not be an easy task. Fortunately, through the work in references \cite{wecker_solving_2015, jiang_quantum_2018}, we have an algorithm to prepare the ground state of this non-interacting fermion Hamiltonian.

The first step is to diagonalize $H_I$ by Fourier (sine) transformation.
Using
% By performing the following unitary transformations on the creation(annihilation) operators defined for the single-particle states localized around the sites of the lattice, we get a new set of creation(annihilation) operators defined for single-particle states that form the eigenbasis of the Hamiltonian in Eq. \eqref{eq:hd},
%
\eql{mom-states}{c_{k\sigma}^\dagger = \sqrt{\frac{2}{L+1}} \sum\limits_{j=1}^L \sin{(kj)} c_{j\sigma}^\dagger\;,}
where $k= {m\pi}/{(L+1)}$; $m=1,2,\dots ,L$,
$H_I$ becomes
\eql{k-h0}{H_I = \sum\limits_{k,\sigma} \epsilon_k c_{k\sigma}^\dagger c_{k\sigma}^{\phantom{\dagger}}\;,}where $\epsilon_k = -2t_H\cos{k}$. 

The Hamiltonian Eq.~\eqref{eq:ha} commutes with the total number operator $N$. Our quantum annealing protocol also preserves the total number of spin-$\uparrow$ electrons $(N_\uparrow)$ and the total number of spin-$\downarrow$ electrons $(N_\downarrow)$. This means that to find the ground state with a specific $N_\uparrow$ and $N_\downarrow$, we need to start in a state with the same $N_\uparrow$ and $N_\downarrow$. 

In fermion language, the ground state of Eq.~\eqref{eq:hd} in a subspace with a given value of $N_\uparrow$ and $N_\downarrow$ reads
\eql{i-gs}{\ket{\psi(0)} = c_{k_1\uparrow}^\dagger c_{k_2\uparrow}^\dagger \dots c_{k_{N_\uparrow}\uparrow}^\dagger \dots c_{k_{N_\downarrow}\downarrow}^\dagger\ket{V}\;,}where $\ket{V}$ is the fermion vacuum state (all zero's) and the $k$'s are chosen so that $\sum \epsilon_k$ is the minimum.
For a simple example, see 
Fig.~\ref{fig:dispersion}. 

Substituting Eq.~\eqref{eq:mom-states} into Eq.~\eqref{eq:i-gs}, we obtain the initial ground state $\ket{\psi(0)}$ in terms of the fermion operators for which we defined the Jordan-Wigner transformation by Eq.~\eqref{eq:jw}. This enables us to write the ground state Eq.~\eqref{eq:i-gs} in the computational basis and subsequently prepare it using Givens rotations as discussed in Appendix~\ref{Givens rotations}.

Being equipped with a circuit that prepares the initial state, we are now in a position to perform quantum annealing starting from this state. The next subsection outlines how we construct the quantum circuit for the quantum annealing process itself.

\subsection{Time-evolution} \label{te}

To simulate quantum annealing, we need to solve the TDSE given by Eq.~\eqref{eq:tdse}. The general solution is given by a unitary operator that transforms the initial state to a state $\ket{\psi(t)}$ at time $t$,
\eql{unitary}{\ket{\psi(t)} = U(t,t_0)\ket{\psi(t_0)}\;.}

If the Hamiltonian $H$ is time-independent, the operator $U$ is given by,
\eql{tise}{U(t,t_0) = e^{-iH(t-t_0)}\;,}where we have set $\hbar =1$.

The Hamiltonian Eq.~\eqref{eq:q-ham} is time-dependent through the time dependence of $s=t/T_A$ where $T_A$ is the total quantum annealing time. However, if we approximate $H(s)$ as piece-wise constant, the time evolution operator can be well approximated by,
\eql{U_te}{U(T_A,0) \approx \prod\limits_{n=1}^{n={T_A}/{\tau}} e^{-i\tau H(n\tau-{\tau}/{2})}\;,}
for a small time step $\tau$ \cite{suzuki_general_1993}.

The task of simulating the Schr{\"o}dinger dynamics has been reduced to calculating the exponential $e^{-i\tau H(n\tau-{\tau}/{2})}$, and a quantum circuit needs to be constructed for this exponential. Constructing a circuit that performs this operation on the state of a quantum computer exactly is not trivial. Moreover, we want circuits that can be realized with 1-qubit and 2-qubit gates since potential quantum computing architectures might be able to implement these. This means that we need a way to decompose the full unitary evolution into unitary gates that operate on at most 2 qubits.

We use the Suzuki-Trotter approximation \cite{trotter_product_1959, suzuki_generalized_1976} in which, by using a small enough $\tau$, the exponential of a Hamiltonian can be well approximated by a product of exponentials of each term $H_j$ in the Hamiltonian. For each small time step $\tau$ \cite{huyghebaert_product_1990, suzuki_general_1993},
\eql{st}{e^{-i\tau \sum\limits_j H_j(n\tau-{\tau}/{2})} \approx \prod\limits_j e^{-i\tau H_j(n\tau-{\tau}/{2})}\;,}which is the first-order Suzuki-Trotter approximation. The approximation can be improved by using the second-order Suzuki-Trotter approximation \cite{suzuki_decomposition_1985, de_raedt_product_1987},
\begin{align}
\begin{split} \label{sopf}
    e^{-i\tau \sum\limits_j H_j(n\tau-{\tau}/{2})} \approx&
    \prod\limits_{j=1}^{l} e^{-i{\tau}/{2} H_j(n\tau-{\tau}/{2})} \\
    &\times
\prod\limits_{k=1}^{l} e^{-i{\tau}/{2} H_{l-k+1}(n\tau-{\tau}/{2})}\;,
\end{split}
\end{align}
where $l$ is the number of terms in the Hamiltonian.

We use Eq.~(\ref{sopf}) to simulate the evolution of a system governed by the annealing Hamiltonian $H(s)$. A key step in this process is identifying and grouping mutually commuting terms within $H(s)$. Doing so, not only mitigates errors arising from non-commutativity but also enables the design of more efficient quantum circuits. Below, we introduce a natural grouping scheme that respects the intrinsic symmetries of $H(s)$. We introduce
\begin{align}
\begin{split}
\label{eq:commute1}
    XY_1 &= \sum\limits_{\substack{i\ \text{odd} \\ i \neq L}}^{2L-1}(X_iX_{i+1}+Y_iY_{i+1}) \;,\\ XY_2 &= \sum\limits_{\substack{i\ \text{even} \\ i \neq L}}^{2L-1} (X_iX_{i+1}+Y_iY_{i+1})\;,\\ ZZ &= \sum\limits_{i=1}^L (I-Z_i)(I-Z_{i+L})\;.
\end{split}
\end{align}

This choice exactly preserves $N_\uparrow$ and $N_\downarrow$, as should be the case with the exact time evolution. 
However, since the time step $\tau$ is chosen to be sufficiently small, $N_\uparrow$ and $N_\downarrow$ will be approximately preserved also with other groupings. 
Therefore, we adopt the alternative grouping, 
\begin{align}
\begin{split}
\label{eq:commute2}
XX &= \sum\limits_{\substack{i=1 \\ i \neq L}}^{2L-1} X_iX_{i+1} \;,\\ YY &= \sum\limits_{\substack{i=1 \\ i \neq L}}^{2L-1} Y_iY_{i+1}\;,\\ ZZ &= \sum\limits_{i=1}^L (I-Z_i)(I-Z_{i+L})\;,
\end{split}
\end{align}
which provides a substantial reduction of the total number of gates over the choice above, facilitating more efficient circuits.
Since for the Pauli matrices $\sigma_i^\alpha$ with direction $\alpha$ that act on qubit $i$, $[\sigma_i^\alpha\sigma_j^\alpha,\sigma_m^\alpha\sigma_n^\alpha]=0$, the operation $e^{-iXX}$, $e^{-iYY}$ and $e^{-iZZ}$ can be decomposed exactly.

We use the following universal gate set to decompose these unitary operators:
\begin{align}
\begin{split}
    &H = \frac{1}{\sqrt{2}}\begin{pmatrix}
     1 & 1 \\ 1 & -1
 \end{pmatrix}, \\
    &\pm X = \frac{1}{\sqrt{2}}\begin{pmatrix}
     1 & \pm i \\ \pm i & 1
 \end{pmatrix}, \\
    &RZ(\theta) = e^{-i\frac{\theta}{2}Z} = \begin{pmatrix}
     e^{-i\frac{\theta}{2}} & 0 \\ 0 & e^{i\frac{\theta}{2}}
 \end{pmatrix}, \\
    &RZZ(\theta) = e^{-i\theta ZZ} = \begin{pmatrix}
     e^{-i\theta} & 0 & 0 & 0 \\0 & e^{i\theta} & 0 & 0 \\
     0&0&e^{i\theta} \\ 0&0&0&e^{-i\theta}
 \end{pmatrix}.
 \end{split}
\end{align}

Now we can write,
\al{eixx}{e^{-i\theta XX} &= \prod\limits_{\substack{i=1 \\ i \neq L}}^{2L-1} e^{-i\theta X_iX_{i+1}}\;,\nonumber \\
&= \prod\limits_{\substack{i=1 \\ i \neq L}}^{2L-1} e^{-i\theta H_iH_{i+1}Z_iZ_{i+1}H_iH_{i+1}}\;, \nonumber\\
&= \prod\limits_{\substack{i=1 \\ i \neq L}}^{2L-1} H_iH_{i+1}e^{-i\theta Z_iZ_{i+1}}H_iH_{i+1}\;, \nonumber\\
&= \prod\limits_{i=1}^{2L}H_i \prod\limits_{\substack{i=1 \\ i \neq L}}^{2L-1} e^{-i\theta Z_iZ_{i+1}}\prod\limits_{i=1}^{2L}H_i\;, \nonumber\\
&= \prod\limits_{i=1}^{2L}H_i \prod\limits_{\substack{i=1 \\ i \neq L}}^{2L-1} RZZ_{i,i+1}(\theta)\prod\limits_{i=1}^{2L}H_i\;.}

Similarly,
\eql{eiyy}{e^{-i\theta YY} = \prod\limits_{i=1}^{2L}(+X_i) \prod\limits_{\substack{i=1 \\ i \neq L}}^{2L-1} RZZ_{i,i+1}(\theta)\prod\limits_{i=1}^{2L}(-X_i)\;.}and,
\al{eizz}{e^{-i\theta ZZ} &= \prod\limits_{i=1}^{2L} e^{i\theta Z_i} \prod\limits_{i=1}^{L} e^{-i\theta Z_iZ_{i+L}}\;,\nonumber \\
&= \prod\limits_{i=1}^{2L} RZ_i(-2\theta) \prod\limits_{i=1}^{L} RZZ_{i,i+L}(\theta)\;.}

For an intermediate point $s$ of the linear annealing schedule, from Eq.~\eqref{sopf}, the time evolution for the piece-wise constant Hamiltonian can then be approximated by, 
\begin{align}
\begin{split}
    &e^{-i\tau \sum\limits_j H_j(s)} \approx \\
    &e^{i\frac{t_H\tau}{4} XX} e^{-i\frac{sU\tau}{8}ZZ} e^{i\frac{t_H\tau}{2} YY} e^{-i\frac{sU\tau}{8}ZZ} e^{i\frac{t_H\tau}{4} XX}\;.
    \end{split}
\end{align}

Then, from Eq.~\eqref{eq:U_te}, the decomposition of the full unitary operator that simulates quantum annealing can be approximated by, 
\begin{align}
\begin{split}
    U(&T_A,0) \approx e^{i\frac{t_H\tau}{4} XX} e^{-i\frac{2U\tau T_A-U\tau^2}{16T_A}ZZ} e^{i\frac{t_H\tau}{2} YY} \\
\times    &e^{-i\frac{2U\tau T_A-U\tau^2}{16T_A}ZZ} e^{i\frac{t_H\tau}{2} XX}
    \dots e^{i\frac{t_H\tau}{2} YY} e^{-i\frac{3U\tau^2}{16T_A}ZZ} \\ \times&e^{i\frac{t_H\tau}{2} XX} 
    e^{-i\frac{U\tau^2}{16T_A}ZZ} e^{i\frac{t_H\tau}{2} YY} e^{-i\frac{U\tau^2}{16T_A}ZZ} e^{i\frac{t_H\tau}{4} XX}\;.
    \end{split}
\end{align}

The number of Suzuki-Trotter steps is $T_A/\tau$ (chosen to be an integer) and each step can be seen to have one $e^{i\theta_1 XX}$ and $e^{i\theta_1 YY}$ operation, and two $e^{-i\theta_2 ZZ}$ operations. The number of 1-qubit gates of one time step is thus $4L+4L+2\times2L=12L$ and the number of 2-qubit gates is $2(L-1)+2(L-1)+2\times L=6L-4$. The total number of gates required to perform quantum annealing for a given $T_A$ is therefore $(T_A/\tau)(18L-4)$. 

In summary, the product-formula algorithm allows us to perform gate-based quantum annealing simulations by a circuit the size of which scales linearly with system size. The amount of resources required to obtain a reasonable estimate of the ground state of the 1-dimensional Hubbard Hamiltonian then depends on how the required total annealing time $T_A(L)$ scales with the problem size. The total number of gates required to find the ground state is then $\mathcal{O}(L\times T_A(L))$. Discovering how $T_A(L)$ grows with system size allows us to gauge how the amount of quantum computer resources scale with the problem size.

\begin{figure}
\centering
\includegraphics[width=\columnwidth]{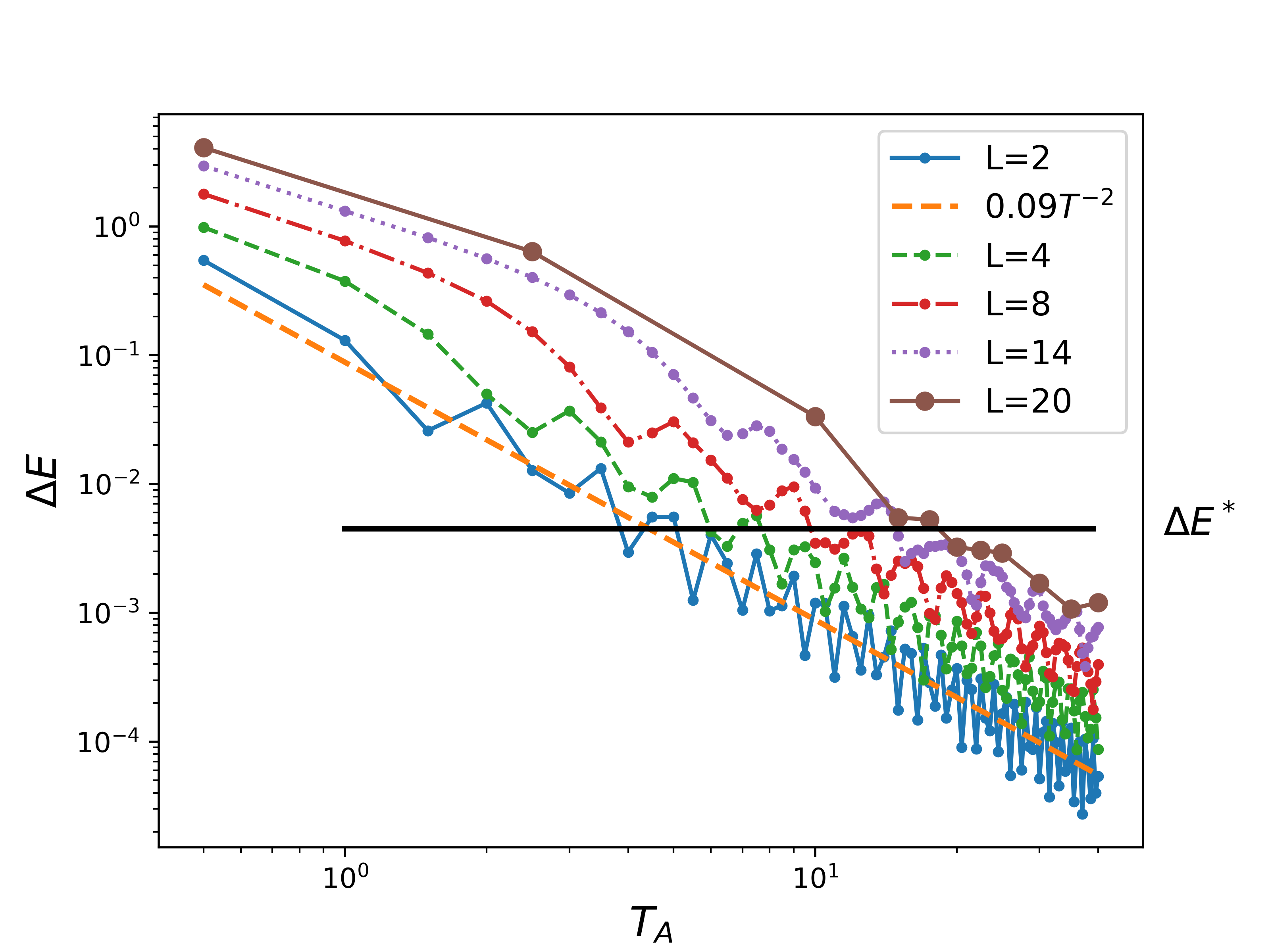}
\caption{\label{fig:residue}$\Delta E$ as a function of $T_A$ for a linear annealing schedule with $U/t_H=4$.}
\end{figure}

\section{results and analysis} \label{results}

The methods discussed in the previous section for performing simulations of gate-based quantum annealing for the problem under consideration can be used to calculate a metric to study the scaling behaviour of quantum annealing to find the ground state of the Hubbard model. For the largest particle number subspaces, corresponding to $2N_\downarrow=N=L$, we use the quantum annealing algorithm to find the ground state of the Hamiltonian Eq.~\eqref{eq:hh}. By doing this for increasing $L$, we infer a scaling of required annealing time from the information of the final quantum state.

\begin{figure*}[t]
    
\begin{subfigure}{\columnwidth}
\centering
\includegraphics[width=\columnwidth]{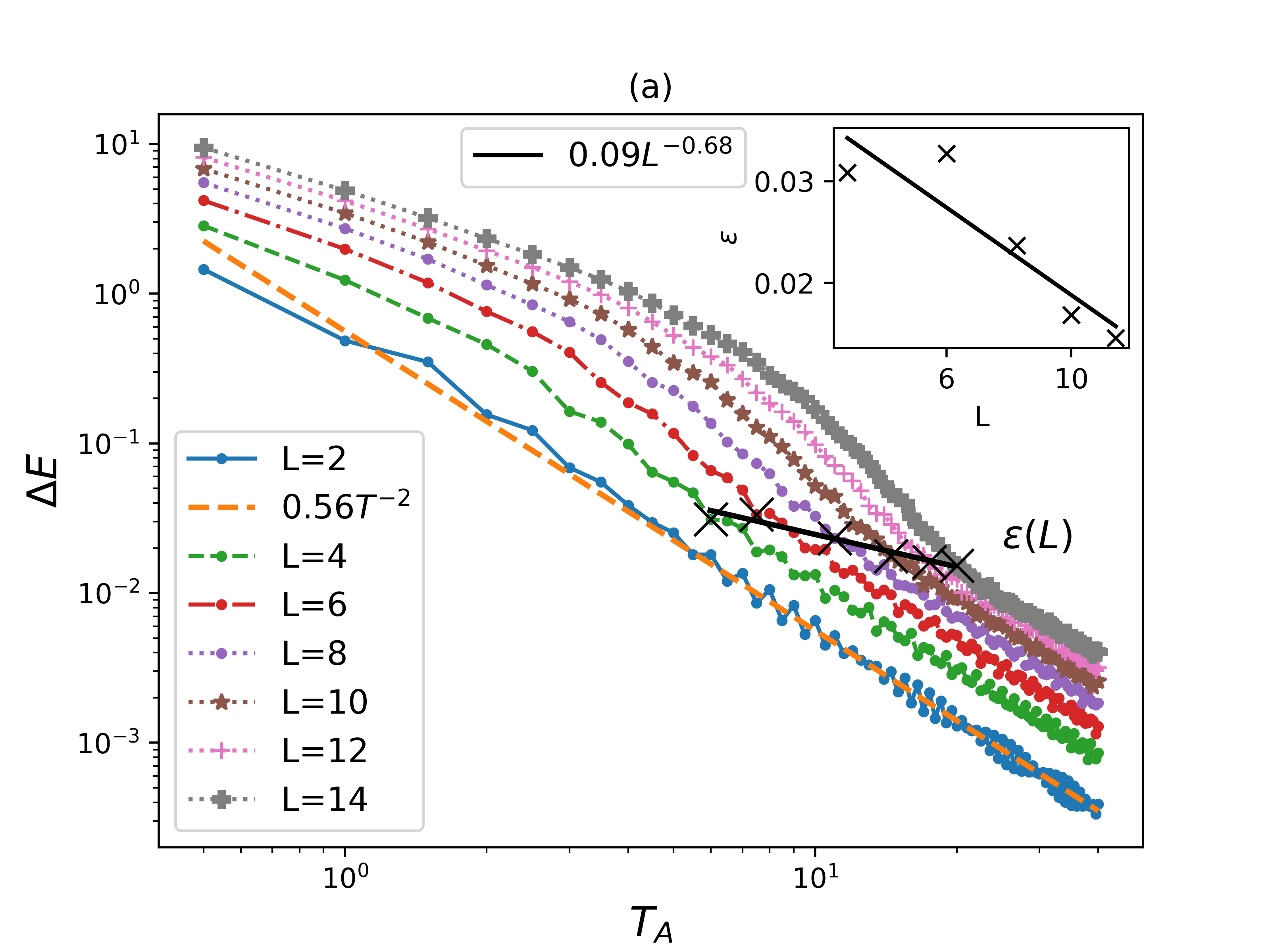}
\captionsetup{labelformat=empty,skip=0pt}
\caption{\mbox{}}
\label{fig:residue_u=8}
\end{subfigure}
\hfill
\begin{subfigure}{\columnwidth}
\centering
\includegraphics[width=\columnwidth]{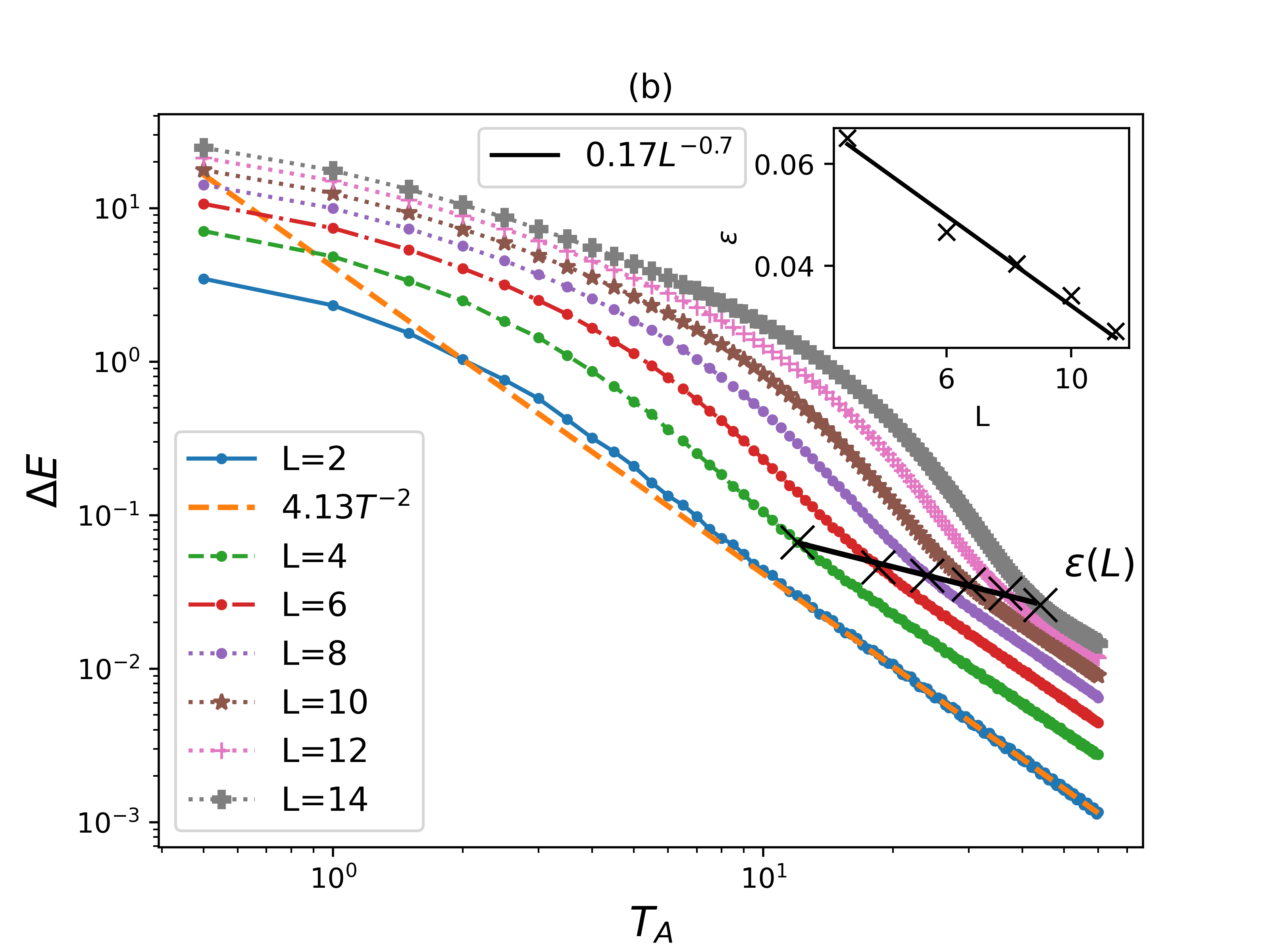}
\captionsetup{labelformat=empty,skip=0pt}
\caption{\mbox{}}
\label{fig:residue_u=16}
\end{subfigure}
\caption{\label{fig:residues_}$\Delta E$ as a function of $T_A$ for a linear annealing schedule with (a) $U/t_H=8$ and (b) $U/t_H=16$. The crosses indicate onsets $\epsilon(L)$, which are plotted as a function of $L$ in the inset.}
\end{figure*}

We construct complete circuits as described in the preceding sections for the simulation of the annealing Hamiltonian Eq.~\eqref{eq:ha}. We do this for systems as large as 20 sites or 40 qubits and for a range of total annealing times $(T_A)$. We use the Jülich Universal Quantum Computer Simulator (JUQCS) \cite{de_raedt_massively_2007, de_raedt_massively_2019}, which is a high-performance program to simulate universal quantum circuits. The simulations were performed on the JUWELS-Booster \cite{kesselheim_juwels_2021} machine at the Jülich Supercomputing Center.

One time step of the quantum annealing algorithm of a 40-qubit system requires 356 gates. For the largest annealing time considered ($T_A=40$) and a time step of $\tau=0.025$, the number of time steps is 1600 and the corresponding number of gates is 569600. Including initial state preparation, the total number of gates in the circuit is then 570918. This problem is simulated on 256 nodes of the JUWELS-Booster using 1024 GPUs and a total computing time of approximately 12 hours.

From the final state $\ket{\psi(s=1)}$ obtained by the simulation, the program calculates the energy $\bra{\psi(1)}H_H\ket{\psi(1)}$. We do this for a range of $T_A$, and we obtain the energy of the final state as a function of total annealing time.

We consider the residual energy as a metric to quantify the proximity of the state obtained after quantum annealing to the exact ground state. Figure~\ref{fig:residue} shows the residue $\Delta E=\bra{\psi(1)}H_H\ket{\psi(1)}-E_0(L)$ versus $T_A$ for a few system sizes. 
The exact ground state energy $E_0(L)$ is obtained
through the numerical solution of the Bethe-ansatz equations,
as outlined in Appendix~\ref{ba_equations}, where we also present the scaling of the required computation time with system size. Table~\ref{tab:simpletable} shows $E_0(L)$ for $L=2$ to $L=20$, $2N_\downarrow=N=L$, and $U/t_H=4,8,16$.
% For $U=4t_H$ we find $E_0(L=2)=??$, $E_0(L=4)=??$, $E_0(L=8)=??$, $E_0(L=14)=??$, and
% $E_0(L=20)=??$.

\begin{table}[ht]
\centering
\begin{tabular}{l@{\hspace{20pt}}c@{\hspace{20pt}}c@{\hspace{20pt}}r}
\hline
% \toprule\toprule
\textbf{$L$} & \textbf{$U=4t_H$} & \textbf{$U=8t_H$} & \textbf{$U=16t_H$} \\
\hline
%\midrule
2 & -0.828427 & -0.472136 & -0.246211\\
%\hline
4 & -1.953145 & -1.117172 & -0.582635\\
%\hline
6 & -3.092565 & -1.768099  & -0.921917\\
%\hline
8 & -4.235807 & -2.420831 & -1.262136\\
%\hline
10 & -5.380619 & -3.074389 & -1.602785\\
%\hline
12 & -6.526243 & -3.728396 & -1.943669\\
%\hline
14 &  -7.67235 & -4.382676 & -2.284694\\
%\hline
16 & -8.818767 & -5.037134 & -2.625814 \\
%\hline
18 & -9.965399 & -5.691715 & -2.966997\\
%\hline
20 & -11.112185 & -6.346385 & -3.308227\\
\hline
%\bottomrule\bottomrule
% \toprule\toprule
\end{tabular}
\caption{Ground state energies $E_0(L)$ for $H_H$ with $2N_\downarrow=N=L$ and $U/t_H=4,8,16$.}
\label{tab:simpletable}
\end{table}

From Fig.~\ref{fig:residue}, it follows that for a certain value of the energy residue which depends on system size, say $\epsilon(L)$, when $\Delta E \leq \epsilon(L)$, the following holds consistently across system sizes:
\eql{residue}{\Delta E \approx \frac{\alpha(L)}{T_A^2}\;.}

From the right hand side of Eq.~\eqref{eq:bound}, we expect that the transition probability should behave like $T_A^{-2}$ for sufficiently long annealing times~\cite{morita_mathematical_2008}. Clearly, the $T_A$-dependence of the energy residue shows a similar behaviour. The adiabatic theorem thus provides support for the observed $T_A^{-2}$ behaviour of the residue and $\epsilon(L)$ marks its onset.

% This means that for,

% \eqn{residue-prec}{\Delta E < \epsilon(L),}we need,
% \eql{precision}{T_A > \sqrt{\frac{\alpha(L)}{\epsilon(L)}}.}

% $\alpha(L)$ and $\epsilon(L)$ will determine how the required annealing time scales to achieve a $\frac{1}{T_A^2}$ behaviour of the energy residue.

We examine the time required by the quantum annealing algorithm to determine the ground state energy with a certain prescribed precision $\Delta E^*$ (see Fig.~\ref{fig:residue}). Then, there is a system size $L^*$ for which the point of onset corresponds to $\epsilon(L^*)\approx\Delta E^*$. For all $L<L^*$, the required annealing time to obtain this precision can be determined from the $L$-dependence of $\alpha(L)$.

% The scaling of $\alpha$ with system size would inform us how the time required for quantum annealing to find the ground state of interest grows with system size for a given precision.

\begin{figure}[h]
\centering
\includegraphics[width=\columnwidth]{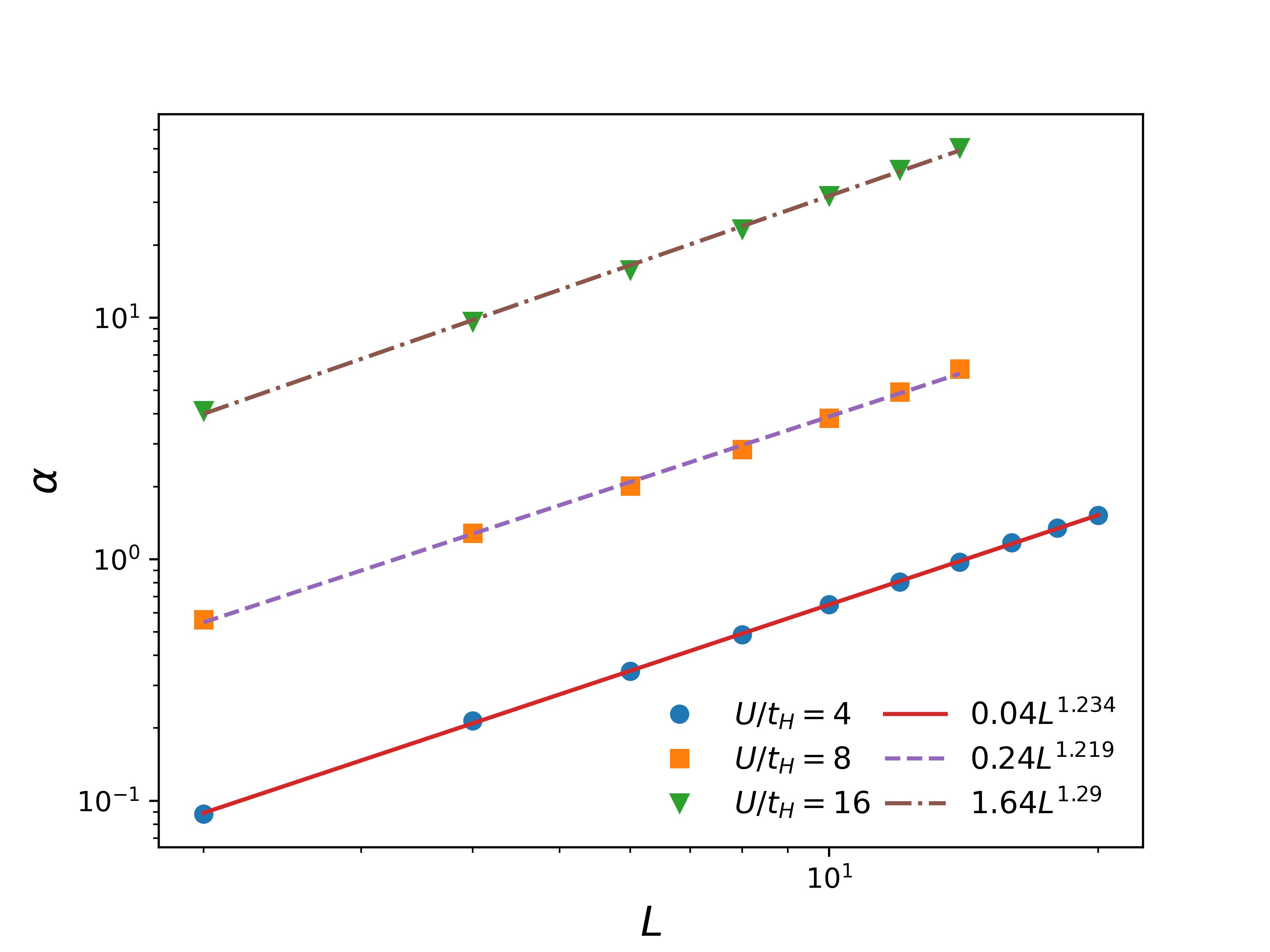}
\caption{\label{fig:advantage}$\alpha(L)$ as a function of $L$ for a linear annealing schedule and $U/t_H=4,8,16$.}
\end{figure}

We plot $\alpha(L)$ with respect to system size and show the result in Fig.~\ref{fig:advantage}. For $U/t_H=4$, the plot fits very well to \eql{alpha}{\alpha(L) \propto L^{1.234}\;.} Substitution in Eq.~\eqref{eq:residue} yields (for $\Delta E \leq \epsilon(L)$),
\begin{align}
\Delta E \propto \frac{L^{1.234}}{T_A^2}\;,    
\end{align}
which means that to obtain the ground state energy with a specified precision $\Delta E^* \approx \epsilon(L^*)$,
\eql{advantage}
{T_A \propto \frac{L^{0.62}}{\sqrt{\Delta E^*}}\;,}where $L<L^*$.

The largest systems up to which Eq.~(\ref{eq:advantage}) holds depends on the scaling of $\epsilon(L)$ with system size. If $\epsilon(L)$ remains constant or decreases polynomially, then the annealing time to obtain the ground state energy with specified precision $\Delta E^*$ only grows polynomially. For a system size $L$ for which $\epsilon(L)<\Delta E^*$, the time required to obtain this precision to the ground state is less than the time required for the onset of the $T_A^{-2}$ regime.

Thus, assuming that 
\eqn{onset_scaling_1}{\alpha(L) \propto L^a \quad \text{and}\quad \epsilon(L) \propto L^{-b}\;,}then for $L>L^*$, where $\Delta E^* > \epsilon(L)$, we have from Eq.~\eqref{eq:residue},
\al{gen_scaling}{&T_A < T_A' \approx \sqrt{\frac{\alpha(L)}{\epsilon(L)}}\propto L^{(a+b)/{2}}\;,}
where $T_A$ is the annealing time to obtain the ground state energy with precision $\Delta E^*$ and $T_A'$ is the annealing time to achieve a $T_A^{-2}$ behaviour of the energy residue.

From the simulation data for $U/t_H=4$, $\epsilon(L)$ cannot be determined with sufficient accuracy. However, Fig.~\ref{fig:residue} already suggests a benign scaling of the onset, which alludes to a polynomial scaling of annealing time for large system sizes. In Appendix \ref{sin schedule}, we show that similar results and analysis for a sinusoidal annealing schedule hold.

To check the performance of quantum annealing in the strong coupling regime we perform simulations for $U/t_H=8,16$. Figs.~\ref{fig:residue_u=8} and \ref{fig:residue_u=16} show the behaviour of the residual energies for $U/t_H=8$ and $U/t_H=16$, respectively. 
The insets show the corresponding estimates for the onsets $\epsilon(L)$ from which we find that 
\eql{onset_scaling_2}{\epsilon(L) \sim L^{-0.7}\;,}providing evidence for the benign scaling of the onset points with system size.

Figure~\ref{fig:advantage} shows the scaling of $\alpha(L)$ for $U/t_H=4,8,16$. The parallel lines are suggestive of a consistent scaling with $L$ dependence.
For $U/t_H=16$,
\eql{alpha3}{\alpha(L) \propto L^{1.29}\;.}
Combining these results to extrapolate the scaling of required $T_A$ to arbitrarily large $L$, we find from Eq.~\eqref{eq:gen_scaling} that,
\eql{final_scaling}{T_A < T_A' \propto L^{0.995}\;.}

To calculate the ground state energy in practice requires performing measurements on the quantum device and sampling the outcomes. Since $H_H$ contains 3 groups of commuting terms, the quantum annealing process needs to be performed $3N_S$ times, where $N_S$ is the number of samples that depends on the desired precision of the measurement. Nevertheless, this does not affect the $L$-dependence of the required total time to calculate the ground state energy, which is $3N_S \times T_A(L)$.

Computing the ground state up to a residual energy of $\Delta E^*$ by quantum annealing, this residue will approximately correspond to $\epsilon(L^*)$ for a certain large $L^*$. From the above arguments, it follows that for system sizes $L<L^*$, quantum annealing yields the ground state of the half-filled one-dimensional Hubbard model in sublinear time. For arbitrarily large $L>L^*$, the required annealing time is bounded by a linear function of system size.

In contrast, although the Bethe-ansatz equations can be solved and the ground state energy can be obtained in polynomial time, determining the ground state requires the calculation of an exponential number of coefficients demanding exponential computational resources. This implies that quantum annealing provides an exponential quantum speed-up over other methods for finding the ground state of the one-dimensional Hubbard model.

\section{Discussion and Outlook} \label{conclusion}

Integrable quantum systems usually have a set of coupled equations that solve the eigenvalue problem of the system exactly. However, to solve these equations, one has to resort to numerical means that require time, which is polynomial in the number of equations. For the one-dimensional Hubbard model with open boundaries, the equations that need to be solved to find the ground state are the Bethe-ansatz Eqs.~\eqref{eq:taka} and \eqref{eq:taka2}. Once solved, the roots can be used to calculate the ground state energy, as well as the ground state coefficients that grow exponentially in number, requiring exponential computer resources. Our analysis of the scaling of required time to find the ground state energy using quantum annealing suggests at least a polynomial quantum speed-up over existing alternatives. Moreover, the speed-up to find the ground state itself is exponential. Even though the problem is \textit{solved} in the thermodynamic limit, demonstrating an exponential improvement using a quantum algorithm for obtaining the ground state reinforces our hope towards the paradigm of quantum computing for solving quantum many-body problems. %This marks a significant step towards understanding the potential of quantum algorithms like quantum annealing in solving quantum many-body systems. 

Quantum annealing is currently being studied extensively for solving various quantum and classical problems \cite{farhi_quantum_2001, brooke_quantum_1999, martonak_quantum_2004, farhi_quantum_2000, munoz-bauza_scaling_2025, somma_quantum_2012, albash_demonstration_2018, mehta_quantum_2022, hsu_quantum_2019, farhi_quantum_2019}. There are problems for which we see promise \cite{albash_demonstration_2018, somma_quantum_2012, munoz-bauza_scaling_2025}, and there are problems for which quantum annealing does not seem to bring any advantage \cite{mehta_quantum_2022}. By approaching quantum many-body problems like the Hubbard model with the algorithm of quantum annealing and investigating its performance, we are able to add significant strength to the promise of the utility of adiabatic quantum computing and quantum annealing.

We find two regimes of scaling depending on the required precision $\Delta E^*$ to find the ground state energy. For system size $L<L^*$, where $L^*$ is the system size for which $\Delta E^*$ corresponds to the onset of the $T_A^{-2}$ behaviour of the residual energy, the obtained time complexity is $\mathcal{O}(L^{0.62})$ and the fits in Fig. \ref{fig:advantage} provide compelling evidence for this scaling. For the $L>L^*$ regime, the scaling estimates are coarse but still suggestive of a bound on the annealing time that is linear in $L$. Nevertheless, $\Delta E^*$ can be chosen to be arbitrarily small, in which case $L^*$ becomes very large. In practice, to obtain a highly precise estimate of the ground state, a sublinear increase of required annealing time would suffice even for large systems. 

Such a scaling behaviour of the required annealing time can be understood to appear as a consequence of the absence of phase transitions in the one-dimensional Hubbard model at nonzero $U$ \cite{lieb_absence_1968, schulz_hubbard_1985}. This means that the dominant gap closing in the spectrum of the annealing Hamiltonian is at $s=0$ \cite{wecker_solving_2015}. Since the gap at $s=0$ decreases linearly, a polynomial scaling of required annealing time is to be expected.

Resource requirements for the considered implementation on a universal quantum computer also show promise for future fault-tolerant architectures. Since the number of gates grows linearly for one Suzuki-Trotter step, the total number of gates for finding a near-precise ground state scales like $\mathcal{O}(L^{1.62})$. Furthermore, the gate depth is constant per Suzuki-Trotter step for the one-dimensional case, which means that the depth also scales like $\mathcal{O}(L^{0.62})$. Even though such computations with thousands of gates might not be a near-term possibility, the scaling of resources augurs well for the era of fault tolerance. On the other hand, analog devices with capabilities of simulating Hubbard-like Hamiltonians \cite{esslinger_fermi-hubbard_2010, hensgens_quantum_2017, tarruell_quantum_2018} might also turn out to be useful in the near future.

Showing evidence of a substantial scaling advantage for an integrable quantum system also invites inquiry into the power of quantum annealing for solving non-integrable quantum many-body systems. However, non-integrable quantum systems, like the two-dimensional Hubbard model, are expected to show the presence of phase transitions. Depending on the order of these phase transitions, the minimum gap in the spectrum of the annealing Hamiltonian can be either exponentially or polynomially decreasing. Considering this, a quantum speed-up for solving the one-dimensional Hubbard model does not necessarily suggest a similar benefit for multi-dimensional many-body quantum systems.

Nonetheless, the question of a quantum advantage using quantum annealing for such systems is still open, and the potential of it persists. Our results drive us to look further into problems like Hubbard models in more dimensions. If not for solving them, quantum annealing can also provide a gateway to understanding phase transitions in these models \cite{wecker_solving_2015}. 

\begin{acknowledgments}
The authors gratefully acknowledge the Gauss Centre for
Supercomputing e.V. for funding this project by providing
computing time through the John von Neumann Institute for
Computing (NIC) on the GCS Supercomputer JUWELS [62]
at Jülich Supercomputing Centre (JSC). This work was supported by the Deutsche Forschungsgemeinschaft (DFG; German Research Foundation) within the Research Unit FOR 2692 under Grant No. 355031190.
\end{acknowledgments}

\appendix

\section{Givens rotations} \label{Givens rotations}
\begin{figure}
    \centering
    \begin{quantikz}
    & \targ{} & \ctrl{1} & \targ{} & \\
    & \ctrl{-1} & \gate{RY(-2\theta)} & \ctrl{-1} &
    % & \gate{H} &
    \end{quantikz}
    \caption{Quantum circuit for implementing a Givens rotation on neighbouring qubits}
    \label{fig:gr_ex}
\end{figure}
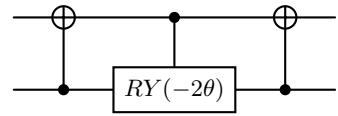
\begin{figure}[tb]
\centering
\includegraphics[width=\columnwidth]{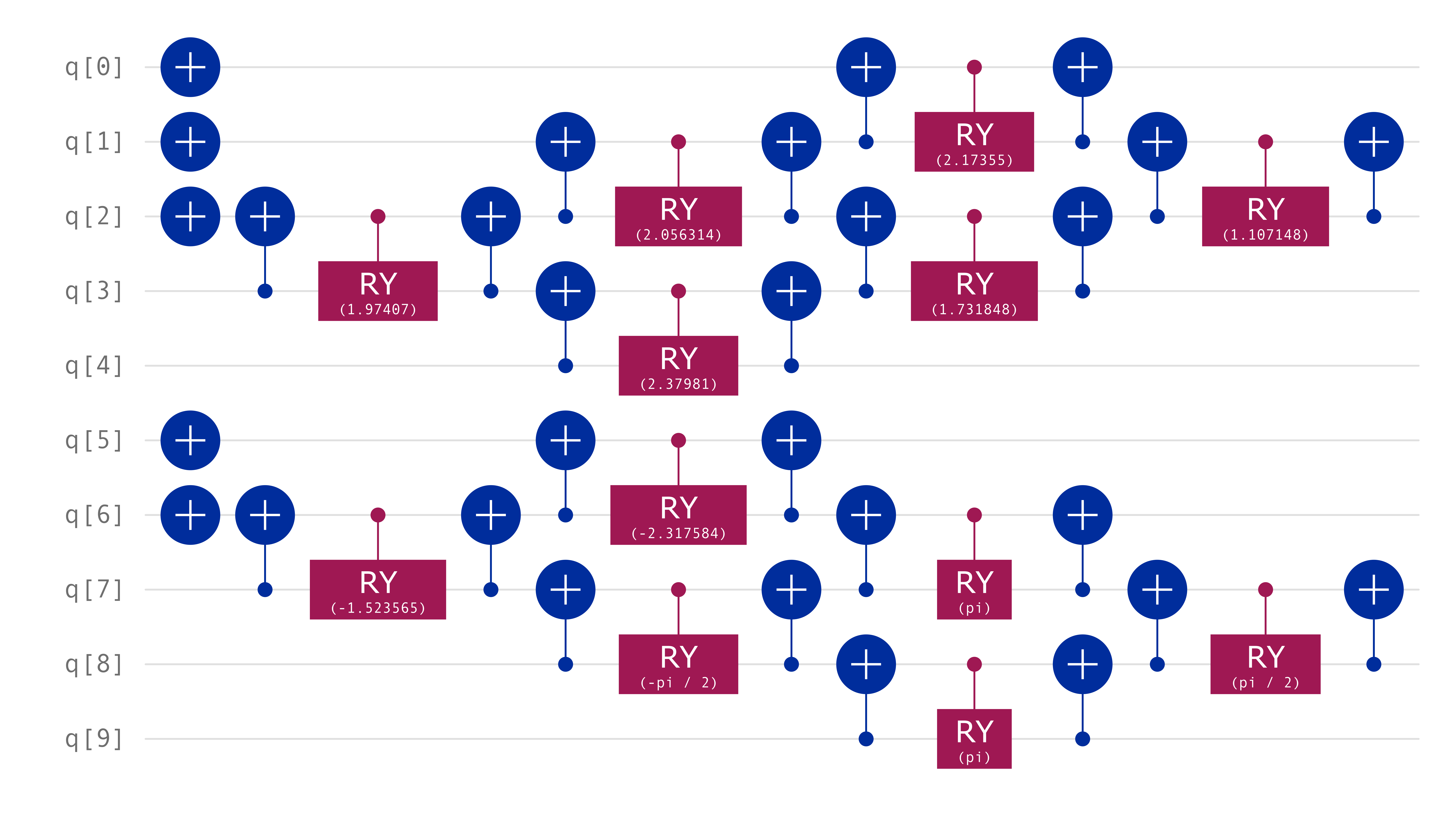}
\caption{\label{fig:givens}Circuit that prepares the ground state of Eq.~\eqref{eq:hd} for $L=5$, $N_\uparrow=3$, $N_\downarrow=2$}
\end{figure}

We follow the method of preparing the initial state using Givens rotations as given in references \cite{wecker_solving_2015, jiang_quantum_2018}.

It can be seen from Eq.~\eqref{eq:mom-states} that the initial state is decoupled in the spin degrees of freedom. This means that we can prepare the state of the first $L$ qubits independently from the state of the last $L$ qubits. Hence, we can suppress the spin index and write a matrix equation like the following:\eql{c-transform}{\begin{pmatrix}
    c_{k_1}^\dagger \\
    \vdots \\
    c_{k_N}^\dagger
\end{pmatrix}=Q\begin{pmatrix}
    c_1^\dagger \\
    \vdots \\
    c_N^\dagger
\end{pmatrix},} where $Q$ is defined by the transformation in Eq.~\eqref{eq:mom-states}. It can be shown that the matrix $Q$ can be decomposed into a sequence of Givens rotations that act on creation operators of neighbouring sites \cite{wecker_solving_2015, jiang_quantum_2018}, where a Givens rotation is defined as,
\eqn{givens}{G_{i,i+1}(\theta)=\begin{pmatrix}
    \cos{\theta} & \sin{\theta} \\
    -\sin{\theta} & \cos{\theta}
\end{pmatrix},}and the angles can be calculated as shown in Ref. \cite{jiang_quantum_2018}.

To see how these Givens rotations that transform the creation operators can be used to prepare the initial state, we can check how they transform a fermion state and map this operation into the one for a qubit state. If we have a fermion state for a system with $L=5$ and $N=3$ as $\ket{11100}$, then
\aln{givens-ex}{G_{3,4}(\theta)&\ket{11100} = G_{3,4}(\theta)c_{1}^\dagger c_{2}^\dagger c_3^\dagger\ket{V}\\
&=\sin{(\theta)}c_{1}^\dagger c_{2}^\dagger c_3^\dagger\ket{V}+\cos{(\theta)}c_{1}^\dagger c_{2}^\dagger c_4^\dagger\ket{V}\\
&=\sin{(\theta)\ket{11100}+\cos{\theta}\ket{11010}}\;.}

Then, a circuit that implements a Givens rotation on two neighbouring qubits can be constructed as shown in Fig. \ref{fig:gr_ex}. Once the sequence of Givens rotations is known, the full circuit that prepares the initial state can be constructed. 
Figure \ref{fig:givens} shows a circuit that prepares the initial state for a system of spin-$\frac{1}{2}$ fermions with $L=5$, $N_\uparrow=3$ and $N_\downarrow=2$.

\section{Solving the Bethe-ansatz equations}\label{ba_equations}

In order to calculate the residual energies for the problems considered, we need to obtain the corresponding values of the exact ground state energies $E_0(L,N,N_\downarrow)$ by solving the Bethe-Ansatz equations. There is no known analytical solution for these coupled algebraic equations due to their nonlinearity. However, they are quite tractable numerically. We use standard software packages that solve nonlinear equations and find the roots of Eqs.~\eqref{eq:taka} and \eqref{eq:taka2}.

\begin{figure}[t!]
\centering
\includegraphics[width=\columnwidth]{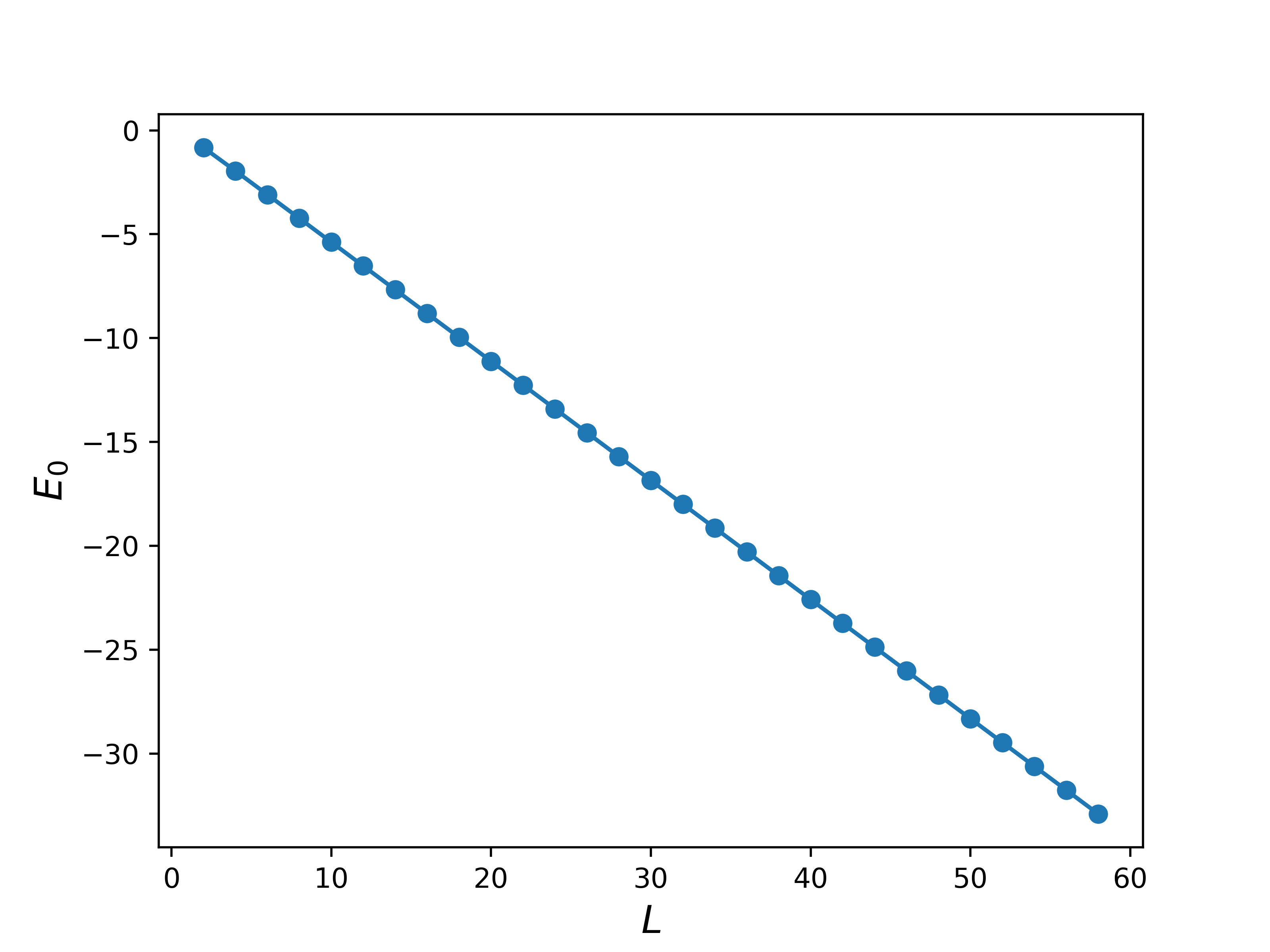}
\caption{\label{fig:exact_energies}$E_0$ as a function of $L$ for Hubbard Hamiltonians with $2N_\downarrow=N=L$ and $U=4t_H$.}
\end{figure}

We are interested in finding the ground state energy of the half-filled Hubbard model ($2N_\downarrow=N=L$) for a range of system sizes $L$. We thus calculate $E_0(L)=E_0(L,N=L,N_\downarrow=L/2)$ using Eq.~\eqref{eq:energy} and the obtained roots $\{ k_1,k_2 \dots k_L \}$. Figure~\ref{fig:exact_energies} shows the $L$-dependence of $E_0(L)$.

\begin{figure}[t!]
\centering
\includegraphics[width=\columnwidth]{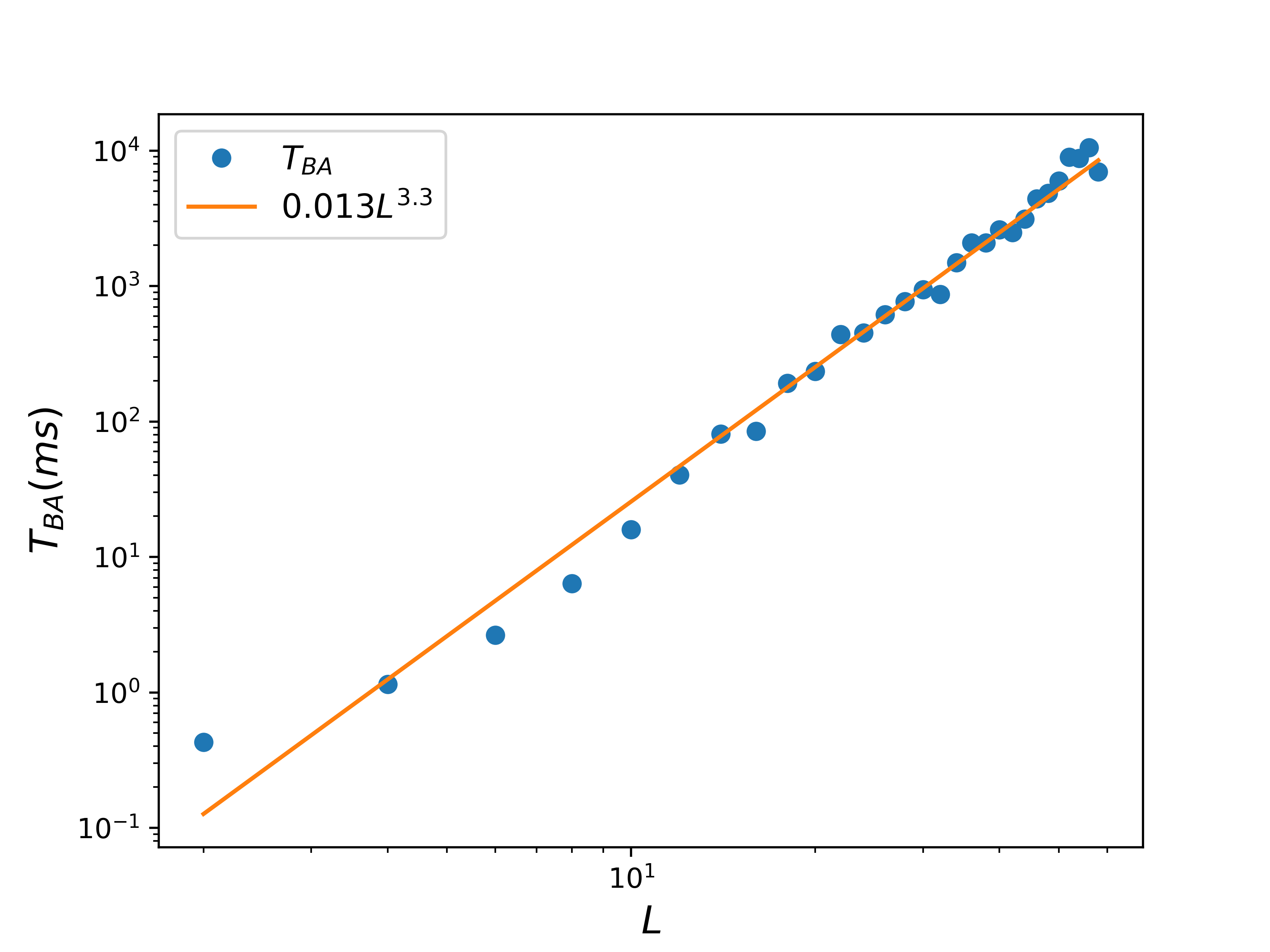}
\caption{\label{fig:exact_energies_time}The computation time $T_{BA}$ to solve the Bethe-ansatz equations as a function of $L$ for Hubbard Hamiltonians with $2N_\downarrow=N=L$ and $U=4t_H$.}
\end{figure}

To get an idea of how the implemented root finding algorithm performs to find the ground state energy, we plot the time the computer program takes to find the roots of the Bethe-ansatz equations for each $L$ in Fig.~\ref{fig:exact_energies_time}. Clearly, the data can be fit to,
\begin{align}
    T_{BA} \propto L^{3.3}
    \;,
\end{align}indicating the polynomial growth of the required time to find the ground state energy by solving the Bethe-ansatz Eqs.~\eqref{eq:taka} and \eqref{eq:taka2}.

\begin{figure*}[t!]
\begin{subfigure}{\columnwidth}
\centering
\includegraphics[width=\columnwidth]{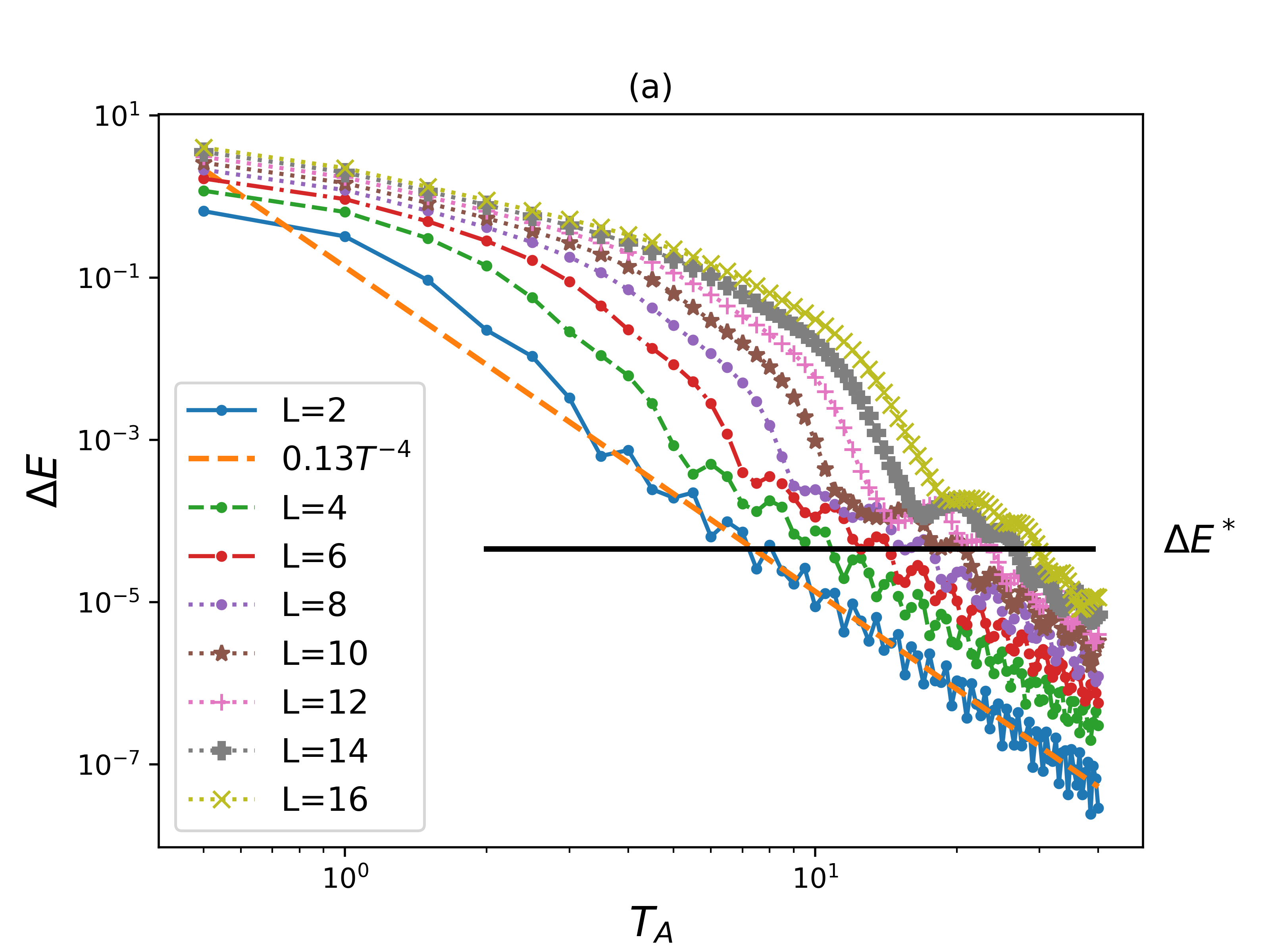}
\captionsetup{labelformat=empty,skip=0pt}
\caption{\mbox{}}
\label{fig:residue_sin}
\end{subfigure}
\hfill
\begin{subfigure}{\columnwidth}
\centering
\includegraphics[width=\columnwidth]{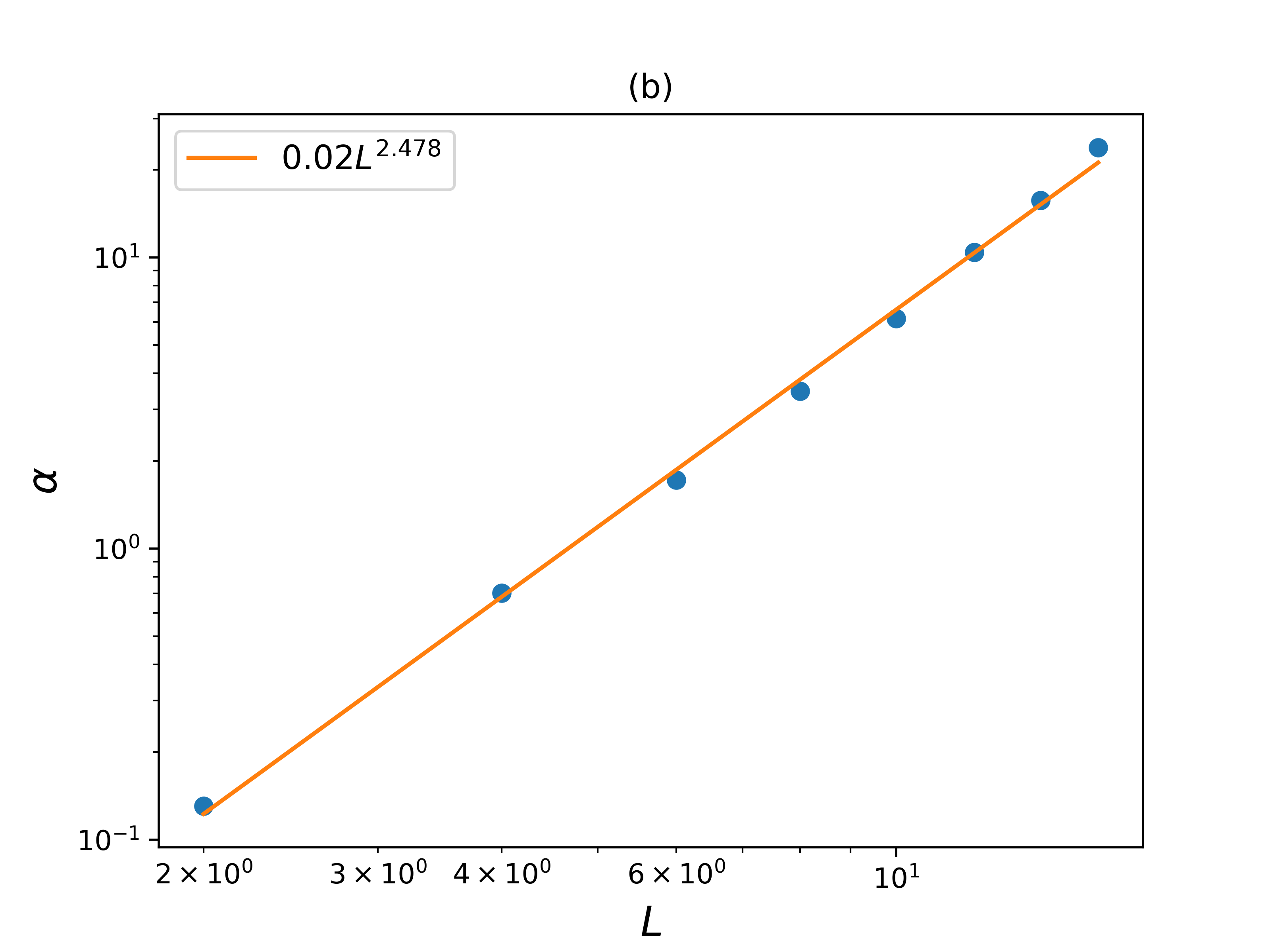}
\captionsetup{labelformat=empty,skip=0pt}
\caption{\mbox{}}
\label{fig:advantage_sin}
\end{subfigure}
\caption{\label{fig:sin_results}(a) $\Delta E$ as a function of $T_A$ and (b) $\alpha(L)$ as a function of $L$ for a sinusoidal annealing schedule with $U/t_H=4$.}
\end{figure*}

\section{Sinusoidal schedule} \label{sin schedule}

We repeat the simulations for the annealing schedule with a sinusoidal dependence on $s$. Namely,
\begin{align}
\begin{split}
H(s) = &-t \sum\limits_{\langle i,j\rangle ,\sigma} ^L(c_{i\sigma}^\dagger c_{j\sigma}^{\phantom{\dagger}} + c_{j\sigma}^\dagger c_{i\sigma}^{\phantom{\dagger}}) \\ &+ \frac{\sin(\pi s-\frac{\pi}{2})+1}{2}U \sum\limits_i^L n_{i\uparrow}n_{i\downarrow}\;.
\end{split}
\end{align}

For such a schedule with $\dot{H}(0)=\dot{H}(1)=0$, the dependence of the bound in Eq.~\eqref{eq:bound} changes to \cite{jansen_bounds_2007, reichardt_quantum_2004, morita_mathematical_2008}
\eql{bound_sin}{1-|\braket{\phi_0(1)|\psi(1)}|^2 \leq \frac{\kappa}{T_A^4}\;,}
where $\kappa$ is independent of $T_A$, and thus, we also expect a $T_A^{-4}$ behavior of $\Delta E$ for large $T_A$. Figure \ref{fig:residue_sin} shows this behavior where we see for a system size dependent onset $\epsilon(L)$, when $\Delta E \leq \epsilon(L)$,
\eql{residue_sin}{\Delta E \approx \frac{\alpha(L)}{T_A^4}\;.}

In Fig. \ref{fig:advantage_sin}, we plot the simulation data for $\alpha(L)$ with $L=2$ to $L=16$. We obtain,
\eql{alpha_sin}{\alpha(L) \propto L^{2.478}\;.}By substituting in Eq.~\eqref{eq:residue_sin}, we find (for $\Delta E \leq \epsilon(L)$)
\eql{residuevL_sin}{\Delta E \propto \frac{L^{2.478}}{T_A^4}\;.}

For a required precision to the ground state energy $\Delta E^* \approx \epsilon(L^*)$, we have,
\eql{advantage_sin}{T_A \propto \frac{L^{0.62}}{\sqrt{\Delta E^*}}\;,}where $L<L^*$.

For this case also, $\epsilon(L)$ could not be determined accurately. However, an indication of a polynomial scaling of the onset is still apparent in Fig. \ref{fig:residue_sin}. For $L<L^*$, we see a sublinear scaling of $T_A$ with system size, with the same exponent for two different annealing schedules.

\bibliographystyle{apsrev4-2}
% \bibliography{references}
%apsrev4-2.bst 2019-01-14 (MD) hand-edited version of apsrev4-1.bst
%Control: key (0)
%Control: author (72) initials jnrlst
%Control: editor formatted (1) identically to author
%Control: production of article title (-1) disabled
%Control: page (0) single
%Control: year (1) truncated
%Control: production of eprint (0) enabled
%

\end{document}